\documentclass[11pt]{article}
\usepackage{jheppub}
\usepackage{amsmath,amssymb,amsthm,bm,bigdelim,multirow}
\usepackage{color}
\usepackage{booktabs}
\usepackage{hyperref}

\newcommand{\kslash}{k\kern-1ex /}
\newcommand{\pslash}{p\kern-1ex /}
\newcommand{\qslash}{q\kern-1ex /}
\newcommand{\lslash}{l\kern-1ex /}
\newcommand{\sslash}{s\kern-1ex /}
\newcommand{\Dslash}{D\kern-1.2ex /}

\newcommand{\beqa}{\begin{eqnarray}}
\newcommand{\eeqa}{\end{eqnarray}}
\newcommand{\be}{\[}
\newcommand{\ee}{\]}
\newcommand{\bd}{\begin{description}}
\newcommand{\ed}{\end{description}}

\newcommand{\ben}{\begin{eqnarray}}
\newcommand{\een}{\end{eqnarray}}

\def\lsim{\raise0.3ex\hbox{$<$\kern-0.75em\raise-1.1ex\hbox{$\sim$}}}
\def\gsim{\raise0.3ex\hbox{$>$\kern-0.75em\raise-1.1ex\hbox{$\sim$}}}
\def\simgt{\rlap{\lower 3.5 pt\hbox{$\mathchar \sim$}}\raise 2.0pt \hbox {$>$}}
\def\simlt{\rlap{\lower 3.5 pt\hbox{$\mathchar \sim$}}\raise 2.0pt \hbox {$<$}}

\begin{document}
  \title{Three-dimensional finite temperature Z$_2$ gauge theory with tensor network scheme}

  \author[a]{Yoshinobu Kuramashi}
  \author[a]{Yusuke Yoshimura}
  \affiliation[a]{Center for Computational Sciences, University of Tsukuba, Tsukuba, Ibaraki
    305-8577, Japan}

  \abstract{
	We apply a tensor network scheme to finite temperature Z$_2$ gauge theory in 2+1 dimensions. 
	Finite size scaling analysis with the spatial extension up to $N_{\sigma}=4096$ at the temporal extension of $N_\tau=2,3,5$ allows us to determine the transition temperature and the critical exponent $\nu$ at high level of precision, which shows the consistency with the Svetitsky-Yaffe conjecture.
  }
\date{\today}

\preprint{UTHEP-723, UTCCS-P-115}

\maketitle

\section{Introduction}
\label{sec:intro}

Tensor network algorithms for coarse-graining of the classical partition functions were originally developed in the field of condensed matter physics \cite{ctmrg,vdm,trg,tefr}. In 2007 the tensor renormalization group (TRG) was proposed by Levin and Nave to study two-dimensional (2D) classical models \cite{trg}. This work attracted the attention of elementary particle physicists so that exploratory studies were performed for 2D scalar models consisting of continuous variables \cite{phi4} with the TRG method. To study wider variety of models in elementary particle physics with the tensor network algorithms, however, it is necessary to develop efficient algorithms for fermion systems and gauge theories. 
Recently one of the authors and his collaborator have successfully applied the Grassmann tensor renormalization group (GTRG) \cite{schwinger,gtrg1,gtrg2} to determine the phase diagram of the one-flavor Schwinger model with and without the $\theta$ term employing the Wilson fermion formulation \cite{schwinger,schwinger-theta,schwinger-phase}, which are the first applications of the GTRG method to lattice gauge theory including fermions in the path integral formalism. The analyses of these models explicitly demonstrated that the GTRG method does not suffer from the sign problem and the complex action problem. Furthermore, direct treatment of the Grassmann numbers in the GTRG method provides us another virtue that the computational cost is comparable to the bosonic case. This work was followed by an investigation of one-flavor 2D Gross-Neveu model with finite chemical potential \cite{njl-fd} and 2D ${\cal N}=1$ Wess-Zumino model \cite{n1-wz}. The GTRG method was also applied to 3D fermionic systems \cite{gtrg_3d,gtrg_gf}. On the other hand, it has been difficult to develop an efficient tensor network algorithm for gauge theories because of its redundancy of gauge degrees of freedom. Although a couple of attempts have been made so far \cite{spinfoam1,spinfoam2,tn-rep,decorated}, the question of their efficiency for lattice gauge theories still remains. 

In this paper we apply the tensor network scheme to a study of 3D finite temperature Z$_2$ gauge theory on the
$V=N_\sigma^2\times N_\tau$
lattice.  This is a good test bed for the first feasibility study of tensor network scheme for lattice gauge theories. Especially the finite temperature phase transition of this model was already investigated in detail by the high-precision Monte Carlo calculation with the aid of the duality to the 3D Ising spin model \cite{z2-mc}. Numerical efficiency of our method is demonstrated by investigating the phase transition with the use of finite size scaling analyses, in which the spatial extension 
$N_\sigma$ is varied up to 4096 at $N_\tau=2,3,5$. It is another virtue of the tensor network scheme that the computational cost is in proportion to the logarithm of the system size so that we can take very large volume regarded as the thermodynamic limit. The results for the transition temperature and the critical exponent $\nu$ are compared to those obtained by the Monte Carlo method \cite{z2-mc}. We also discuss the consistency with the Svetitsky-Yaffe conjecture \cite{SY-conjecture}.

This paper is organized as follows. In Sec.~\ref{sec:TNscheme} we explain a tensor network scheme and associated numerical algorithm to treat the 3D Z$_2$ gauge theory. Numerical results for the finite temperature phase transition of 3D Z$_2$ gauge theory is presented in Sec.~\ref{sec:num}. Sec.~\ref{sec:concl} is devoted to summary and outlook.

\section{Tensor network scheme}
\label{sec:TNscheme}

\subsection{Tensor network formulation}
\label{subsec:formulation}

The partition function of three-dimensional Z$_2$ gauge theory is given by
\begin{align}
	Z =2^{-3V}\sum_{\{\sigma=\pm 1\}} \prod_{n,\mu>\nu} e^{-\beta\sigma_{n,\mu\nu}}
	\label{eq:pf}
\end{align}
with
\begin{align}
	\sigma_{n,\mu\nu}
	=\sigma_{n,\mu} \sigma_{n+\hat\mu,\nu} \sigma_{n+\hat\nu,\mu} \sigma_{n,\nu},
\end{align}
where $\sigma_{n,\mu}\in \{-1,1\}$ is defined on the link labeled by a site $n=(n_0,n_1,n_2)$ with a direction $\mu=0,1,2$.
${\hat \mu}$ and ${\hat \nu}$ denote unit vectors in $\mu$ and $\nu$ directions, respectively.

\begin{figure}
	\centering
	\includegraphics[scale=0.5]{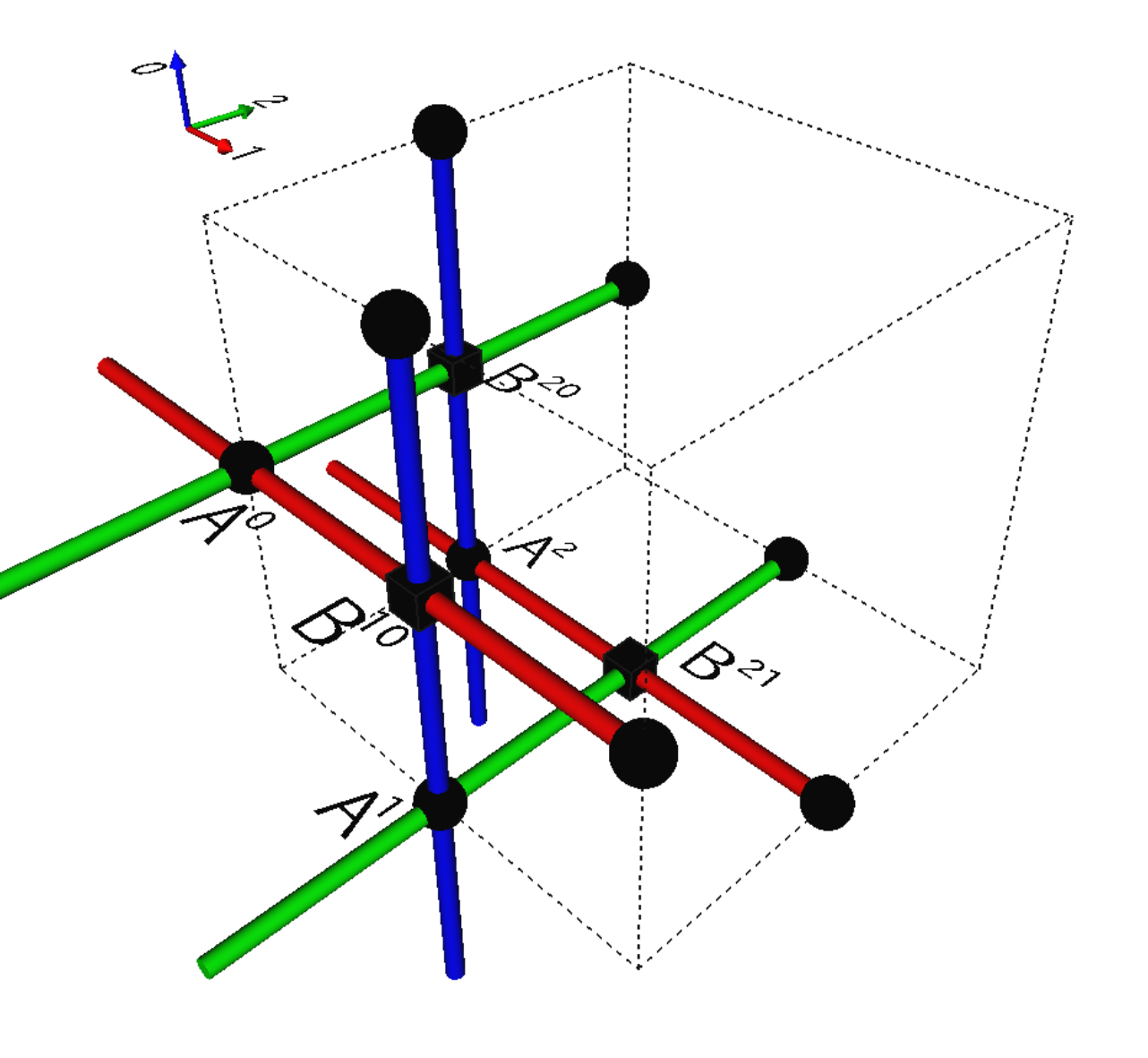}
	\caption{
		Tensor network for Eq~.(\ref{eq:TNF1}).
		Dotted lines describe the cubic lattice.
		Black cubes and spheres represent $B$ and $A$, respectively.
Red, blue and green bars denote the contractions of the tensor indices.  
	}
	\label{fig:TN1}
\end{figure}

We first construct the tensor network representation of Eq.~(\ref{eq:pf}) following Ref.~\cite{tn-rep}.
Each Boltzmann factor at site $n$ is expanded by
\begin{align}
	&e^{\beta \sigma_{n,\mu\nu}}
	=\cosh\beta \sum_{p=0,1} (\tanh\beta)^p \sigma_{n,\mu\nu}^p \nonumber\\
	&\quad
	=\cosh\beta \sum_{p_{n,\mu,\nu},q_{n,\nu,\mu},q_{n,\mu,\nu},p_{n,\nu,\mu}=0,1}
	B^{(n,\mu\nu)}_{p_{n,\mu,\nu}q_{n,\nu,\mu}q_{n,\mu,\nu}p_{n,\nu,\mu}}
	\sigma_{n,\mu}^{p_{n,\mu,\nu}} \sigma_{n+\hat\mu,\nu}^{q_{n,\nu,\mu}}
	\sigma_{n+\hat\nu,\mu}^{q_{n,\mu,\nu}} \sigma_{n,\nu}^{p_{n,\nu,\mu}},
\end{align}
where we introduce the tensor $B$ by
\begin{align}
	B_{pqrs}
	=(\tanh\beta)^{(p+q+r+s)/4} \delta_{p,q}\delta_{q,r}\delta_{r,s}.
\end{align}
We can collect the terms involving a $\sigma_{n,\mu}$ and sum over it.
For example, the case of $\sigma_{n,0}$ is expressed as
\begin{align}
	\sum_{\{\sigma_{n,0}=\pm 1\}} \sigma_{n,0}^{p_{n,0,1}+ p_{n,0,2}+q_{n-\hat 1,0,1}+q_{n-\hat 2,0,2}}
	=2A^{(n,\mu=0)}_{p_{n,0,1}p_{n,0,2}q_{n-\hat 1,0,1}q_{n-\hat 2,0,2}},
\end{align}
where
\begin{align}
	A_{pqrs}= \delta_{\bmod(p+q+r+s,2)=0}.
\end{align}
So the partition function can be rewritten as the following tensor network representation:
\begin{align}
	Z =(\cosh\beta)^{3V} \sum_{\{p,q\}} \prod_{n,\mu>\nu}B^{(n,\mu\nu)} \prod_{n,\mu} A^{(n,\mu)}.
	\label{eq:TNF1}
\end{align}
The connections between tensors are depicted in Fig.~\ref{fig:TN1}.

\begin{figure}
	\centering
	\includegraphics[scale=0.8]{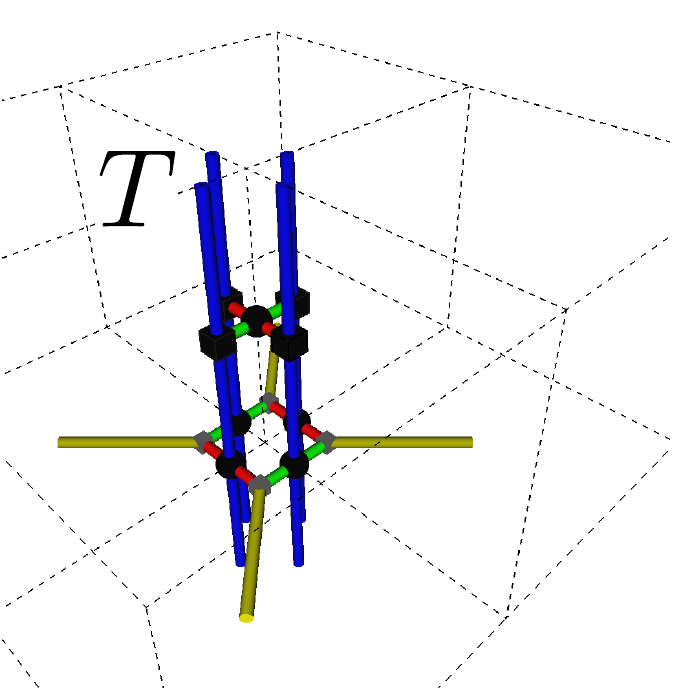}\\
\vspace*{2mm}
	\includegraphics[scale=0.8]{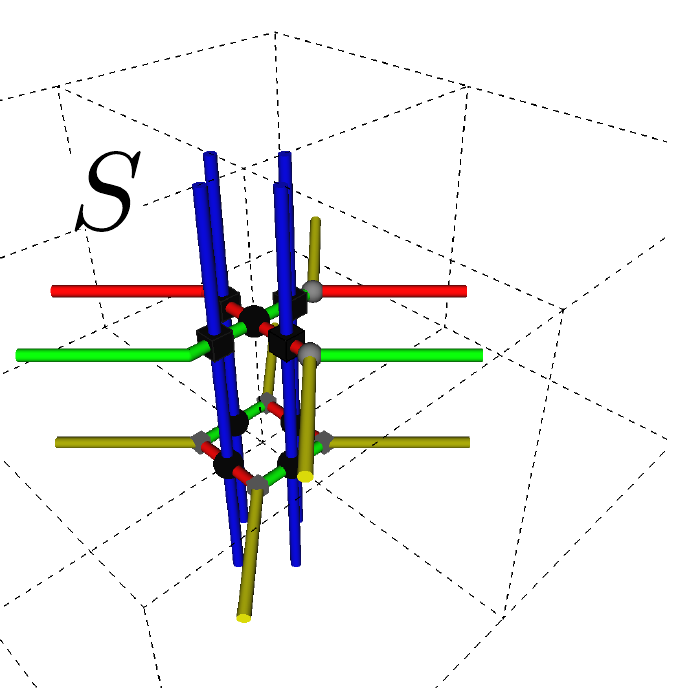}
	\caption{Tensors $T$ (top) and $S$ (bottom) in Eq.~(\ref{eq:Z_TS}).}
	\label{fig:TN_TS}
\end{figure}

In addition, we reconstruct the tensor network representation to reduce the redundant degrees of freedom.
Taking $\mu=0$ as the temporal direction and $\mu=1,2$ as the spatial ones,
we classify the sites as spatially even sites $l$ with
\begin{align}
	\bmod(l_1+l_2,2)=0
\end{align}
and spatially odd sites $m$ with
\begin{align}
	\bmod(m_1+m_2,2)=1.
\end{align}
Let us rearrange the tensors so that all the tensors are collected at 
the spatially even sites.
In this procedure, we employ the gauge fixing with $\sigma_{m,0}=1$ at the spatially odd sites $m$($m_0\ne 0$).
{
Then, all the components of $A^{(m,0)}$ are one constantly and so we can omit the $A^{(m,0)}$ i.e. sum up independently the indecies $p_{m,0,1},p_{m,0,2},q_{m-\hat 1,0,1}$ and $q_{m-\hat 2,0,2}$.
At the $m_0 = 0$ sites, $A^{(m,0)}$ are decomposed as
\begin{gather}
	A^{(m,0)}_{p_{m,0,1}p_{m,0,2}q_{m-\hat 1,0,1}q_{m-\hat 2,0,2}}
	=\sum_{i=0,1} \bar A^{(m-\hat 1)}_{q_{m-\hat 1,0,1}p_{m,0,2}i}
	\bar A^{(m-\hat 2)}_{q_{m-\hat 2,0,2}p_{m,0,1}i}, \\
	\bar A_{pqi}
	=\delta_{\bmod(p+q+i,2)=0}
\end{gather}
as illustrated in Fig.~\ref{fig:TNA}.
\begin{figure}
	\centering
	\includegraphics[scale=0.8]{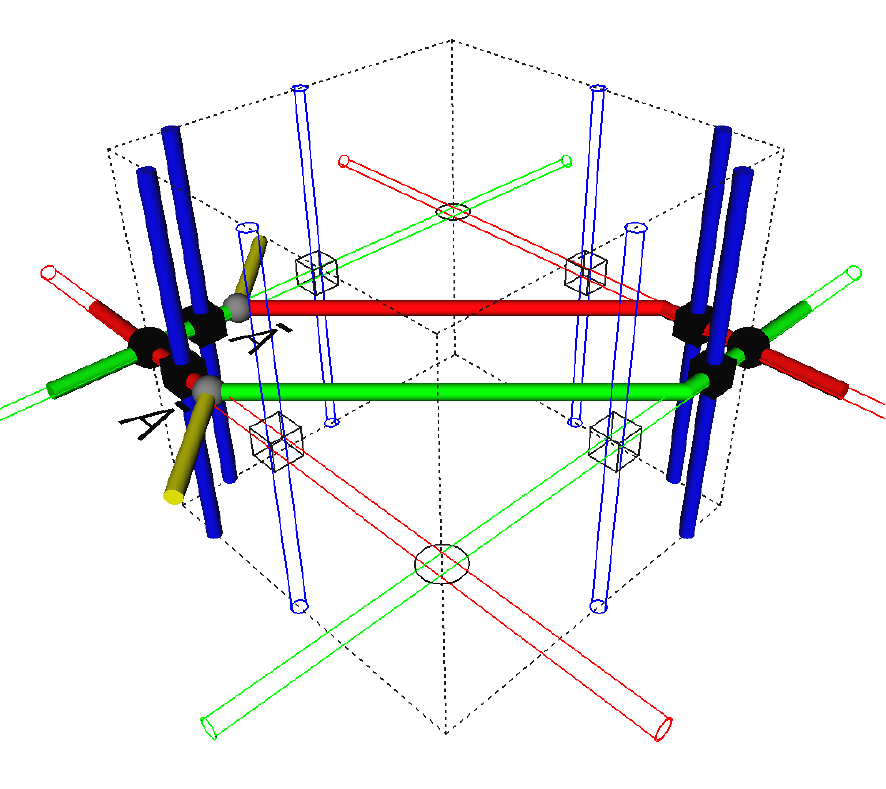}
	\caption{
		Decomposition of $A^{(m,0)}$ at the $m_0 = 0$ sites.
		Gray spheres represent $\bar A$.
	}
	\label{fig:TNA}
\end{figure}
On the other hand, $B^{(n,21)}$ are decomposed as
\begin{gather}
	B^{(l,21)}_{p_{l,2,1}q_{l,1,2}q_{l,2,1}p_{l,1,2}}
	=\sum_{i=0,1}
	\bar B^{(l)}_{p_{l,1,2}p_{l,2,1}i}\bar B^{(l+\hat 1+\hat 2)}_{q_{l,1,2}q_{l,2,1}i}, \\
	B^{(m,21)}_{p_{m,2,1}q_{m,1,2}q_{m,2,1}p_{m,1,2}}
	=\sum_{i=0,1}
	\bar B^{(m+\hat 2)}_{p_{m,2,1}q_{m,1,2}i}\bar B^{(m+\hat 1)}_{q_{m,2,1}p_{m,1,2}i}, \\
	\bar B_{pqi}
	=(\tanh\beta)^{(p+q)/4} \delta_{p,q}\delta_{q,i}
\end{gather}
as illustrated in Fig.~\ref{fig:TNB}.
\begin{figure}
	\centering
	\includegraphics[scale=0.8]{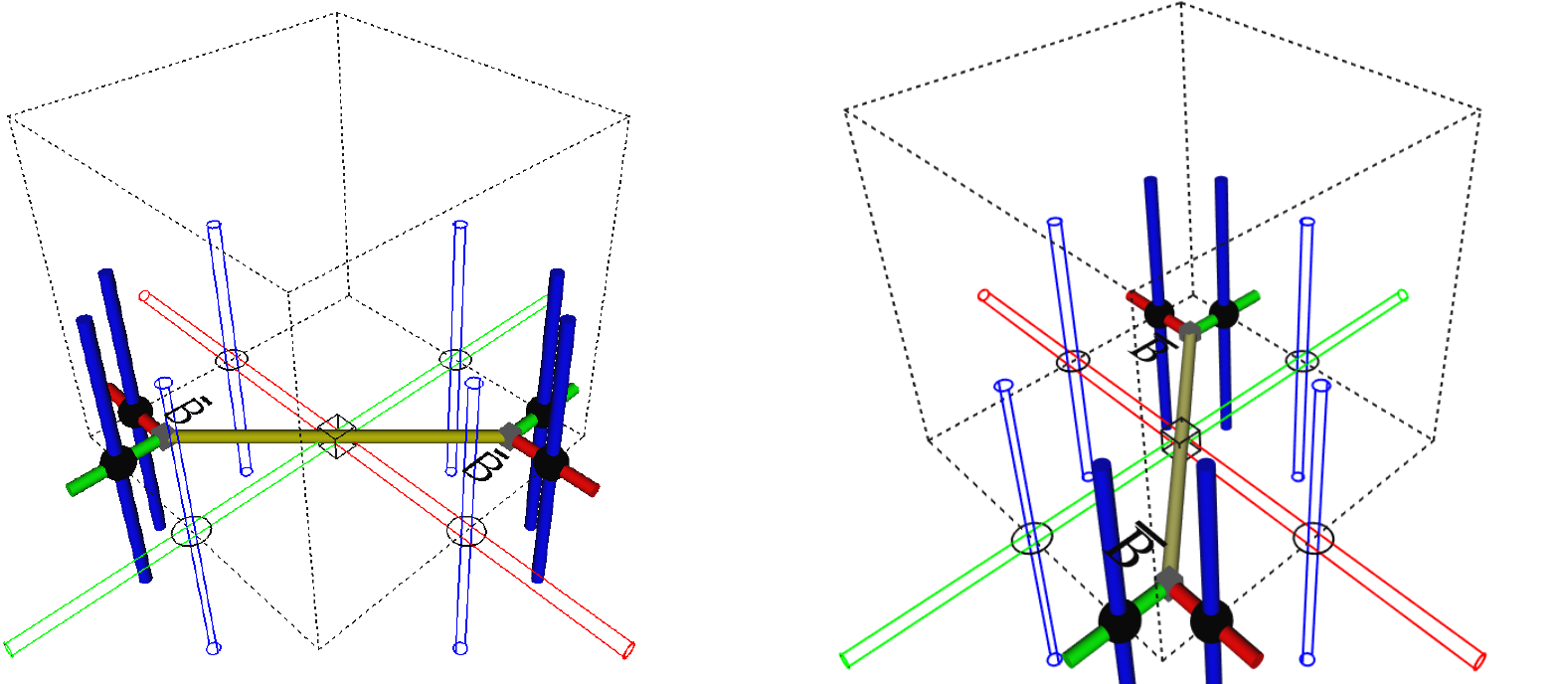}
	\caption{
		Decomposition of $B^{(l,21)}$ (left) and $B^{(m,21)}$ (right).
		Gray cubes represent $\bar B$.
	}
	\label{fig:TNB}
\end{figure}
After contracting all the inner indices of the collected tensors, we obtain new tensors $T$ on the sites of $l_0>0$ as 
\begin{align}
	&T^{(l)}_{xyzx'y'z'}
	=\sum_{p,q} A^{(l,0)}_{p_{l,0,1}p_{l,0,2}q_{l-\hat 1,0,1}q_{l-\hat 2,0,2}}
	\nonumber\\&\quad\cdot
	B^{(l,10)}_{p_{l,1,0}q_{l,0,1}z_1p_{l,0,1}}
	B^{(l,20)}_{p_{l,2,0}q_{l,0,2}z_2p_{l,0,2}}
	B^{(l-\hat 1,10)}_{p_{l-\hat 1,1,0}q_{l-\hat 1,0,1}z_3p_{l-\hat 1,0,1}}
	B^{(l-\hat 2,20)}_{p_{l-\hat 2,2,0}q_{l-\hat 2,0,2}z_4p_{l-\hat 2,2,0}}
	\nonumber\\&\quad\cdot
	A^{(l,1)}_{p_{l,1,2}p_{l,1,0}q_{l-\hat 2,1,2}z'_1}
	A^{(l,2)}_{p_{l,2,0}p_{l,2,1}z'_2q_{l-\hat 1,2,1}}
	A^{(l-\hat 1,1)}_{p_{l-\hat 1,1,2}p_{l-\hat 1,1,0}q_{l-\hat 1-\hat 2,1,2}z'_3}
	A^{(l-\hat 2,2)}_{p_{l-\hat 2,2,0}p_{l-\hat 2,2,1}z'_4q_{l-\hat 2-\hat 1,2,1}}
	\nonumber\\&\quad\cdot
	\bar B^{(l)}_{p_{l,1,2}p_{l,2,1}x}
	\bar B^{(l)}_{q_{l-\hat 1,2,1}p_{l-\hat 1,1,2}y}
	\bar B^{(l)}_{q_{l-\hat 1-\hat 2,1,2}q_{l-\hat 1-\hat 2,2,1}x'}
	\bar B^{(l)}_{p_{l-\hat 2,2,1}q_{l-\hat 2,1,2}y'}
\end{align}
and $S$ on the sites of $l_0=0$ as
\begin{align}
	&S^{(l)}_{xyzx'y'z'}
	=\sum_{p,q} A^{(l,0)}_{p_{l,0,1}p_{l,0,2}q_{l-\hat 1,0,1}q_{l-\hat 2,0,2}}
	\bar A^{(l)}_{q_{l,0,1}x_2y'_2}
	\bar A^{(l)}_{q_{l,0,2}x_3y_2}
	\nonumber\\&\quad\cdot
	B^{(l,10)}_{p_{l,1,0}q_{l,0,1}z_1p_{l,0,1}}
	B^{(l,20)}_{p_{l,2,0}q_{l,0,2}z_2p_{l,0,2}}
	B^{(l-\hat 1,10)}_{p_{l-\hat 1,1,0}q_{l-\hat 1,0,1}z_3x'_3}
	B^{(l-\hat 2,20)}_{p_{l-\hat 2,2,0}q_{l-\hat 2,0,2}z_4x'_2}
	\nonumber\\&\quad\cdot
	A^{(l,1)}_{p_{l,1,2}p_{l,1,0}q_{l-\hat 2,1,2}z'_1}
	A^{(l,2)}_{p_{l,2,0}p_{l,2,1}z'_2q_{l-\hat 1,2,1}}
	A^{(l-\hat 1,1)}_{p_{l-\hat 1,1,2}p_{l-\hat 1,1,0}q_{l-\hat 1-\hat 2,1,2}z'_3}
	A^{(l-\hat 2,2)}_{p_{l-\hat 2,2,0}p_{l-\hat 2,2,1}z'_4q_{l-\hat 2-\hat 1,2,1}}
	\nonumber\\&\quad\cdot
	\bar B^{(l)}_{p_{l,1,2}p_{l,2,1}x_1}
	\bar B^{(l)}_{q_{l-\hat 1,2,1}p_{l-\hat 1,1,2}y_1}
	\bar B^{(l)}_{q_{l-\hat 1-\hat 2,1,2}q_{l-\hat 1-\hat 2,2,1}x'_1}
	\bar B^{(l)}_{p_{l-\hat 2,2,1}q_{l-\hat 2,1,2}y'_1}
\end{align}
as illustrated in Fig.~\ref{fig:TN_TS}.
}
Redefining the spatial even sites $l$ as the new sites $n$, we obtain the final form of the tensor network representation:
\begin{align}
	Z=\sum_{\{t,x,y\}}
	\left( \prod_{n;n_0>0} T^{(n)} \right)
	\left( \prod_{n;n_0=0} S^{(n)} \right),
\label{eq:Z_TS}
\end{align}
where $t$ $(x,y)$ denotes the temporal (spatial) indices in the new network.
Note that the spatial size of the new network is half of the original one.

\subsection{Algorithm for coarse graining}
\label{subsec:algorithm}

\begin{figure}[h]
	\centering
	\includegraphics[scale=0.5]{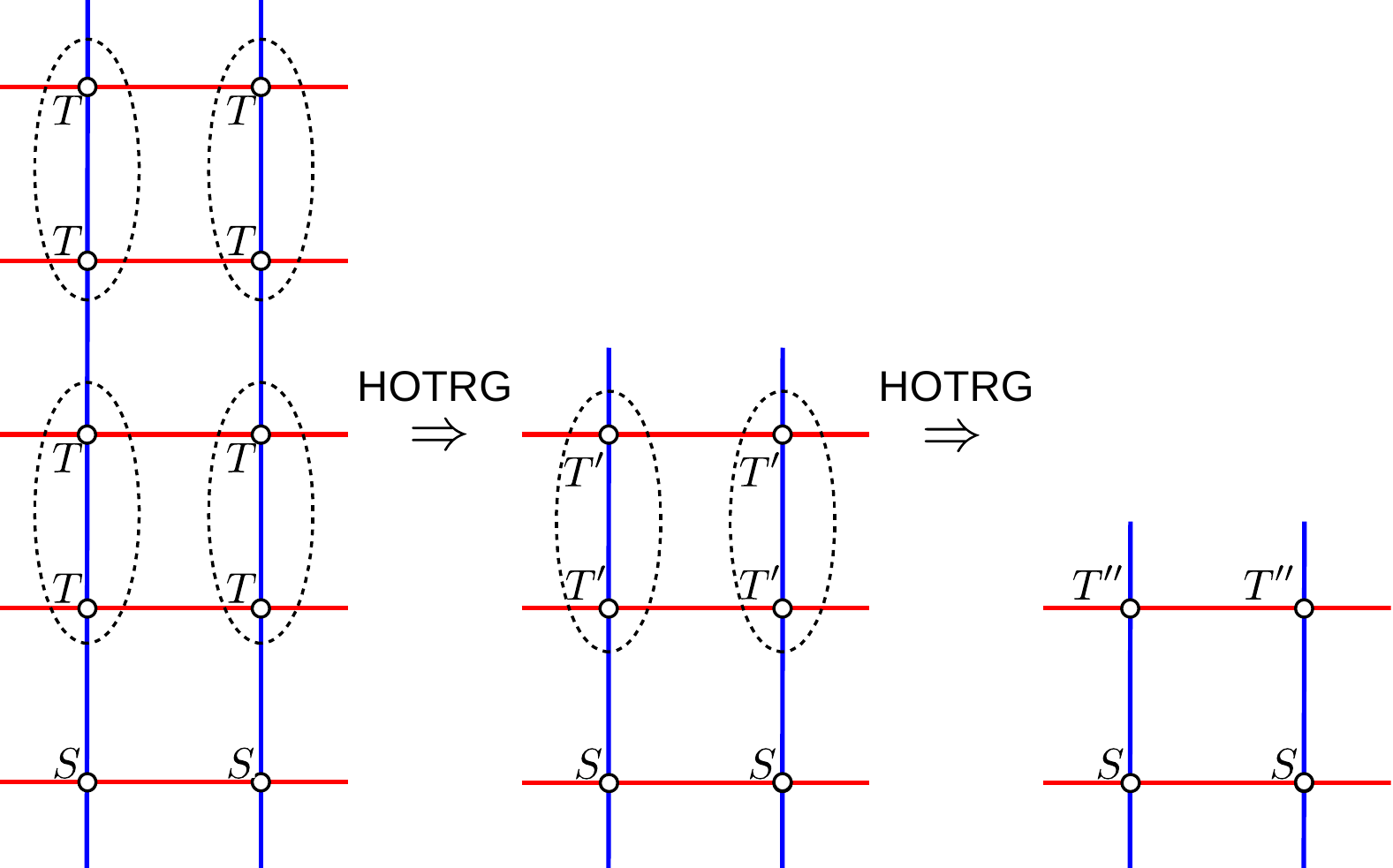}
	\caption{
		The first step of the algorithm.
		Blue (red) line denotes the temporal (spatial) direction.
		After two iterations of the HOTRG procedure, the temporal size of the tensor $T$ is reduced to be one. 
	}
	\label{fig:RA1}
\end{figure}

Our algorithm consists of three steps.
Firstly we repeat coarse-graining of the tensor $T$ in the temporal direction using the HOTRG method \cite{hotrg} until the temporal size of $T$ is reduced to be one. 
In Fig.~\ref{fig:RA1} we illustrate the procedure for the case of $N_\tau=5$.
The dimension of the new tensors $T^\prime$ and  $T^{\prime\prime}$ is truncated to $D_1=16$.
    
\begin{figure}[h]
	\centering
	\includegraphics[scale=0.5]{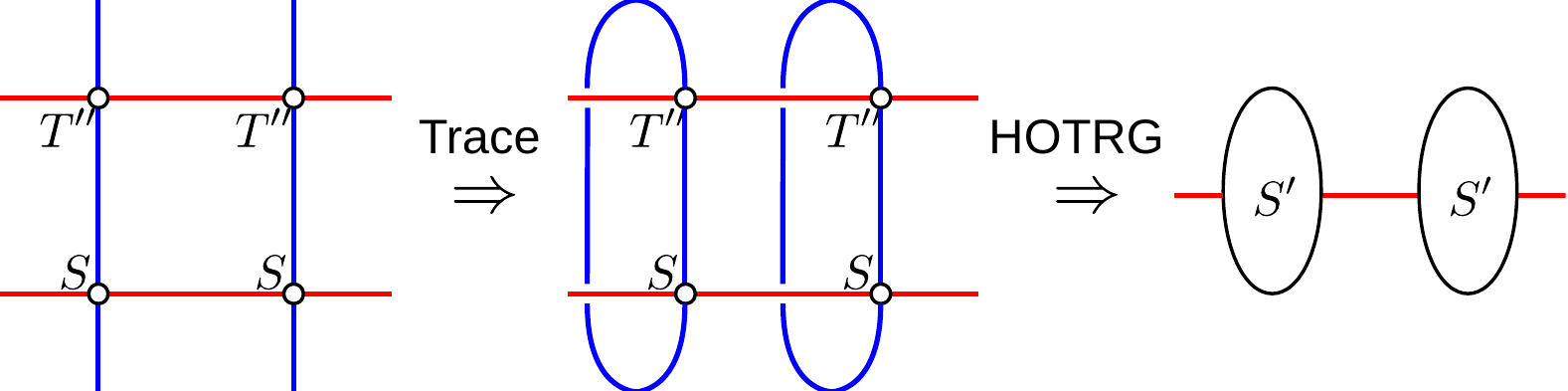}
	\caption{
		The second step of the algorithm. Blue (red) line denotes the temporal (spatial) direction. 
	}
	\label{fig:RA2}
\end{figure}

Secondly we apply the HOTRG procedure to the combination of $T$ and $S$ making the trace of their temporal indices as depicted in Fig.~\ref{fig:RA2}.
The dimension of the new tensor $S^\prime$ is truncated to $D_1=16$.
After this procedure we are left with the two-dimensional tensor network system.
     
Thirdly we make coarse-graining of the two-dimensional system 
until the size is reduced to be $2\times 2$
where the contraction of the indices are exactly carried out.
In this work we employ the TRG method for coarse-graining with $D_2$ as the truncation parameter of the tensor dimensions~{ \footnote{ We need to increase $D_1$ and $D_2$ if we take much larger $N_\tau$}}.

\section{Numerical study}
\label{sec:num}

\subsection{Setup}
\label{subsec:setup}

We determine the phase transition temperature $\beta_c$ and the critical exponent $\nu$ applying the finite size scaling analyses to the specific heat:
\ben
C(N_{\sigma})\equiv \beta^2\frac{\partial^2 \ln Z}{V\partial \beta^2}.
\label{eq:specificheat}
\een
The values of $\beta_c$ and $\nu$ are compared with the previous high-precision Monte Carlo results~\cite{z2-mc}. 

\begin{figure}[h]
    \centering
    \includegraphics[width=80mm]{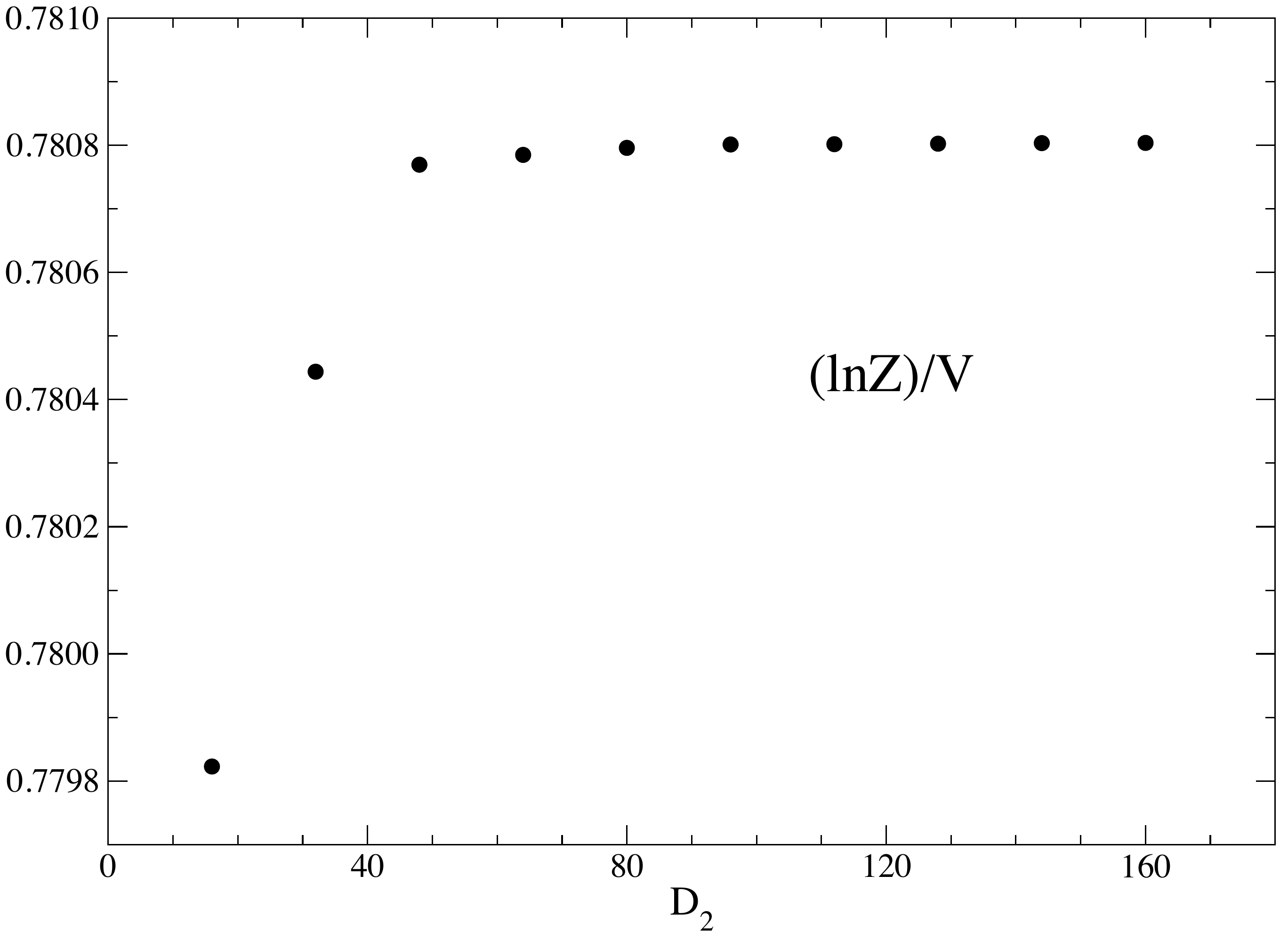}\\
    \includegraphics[width=80mm]{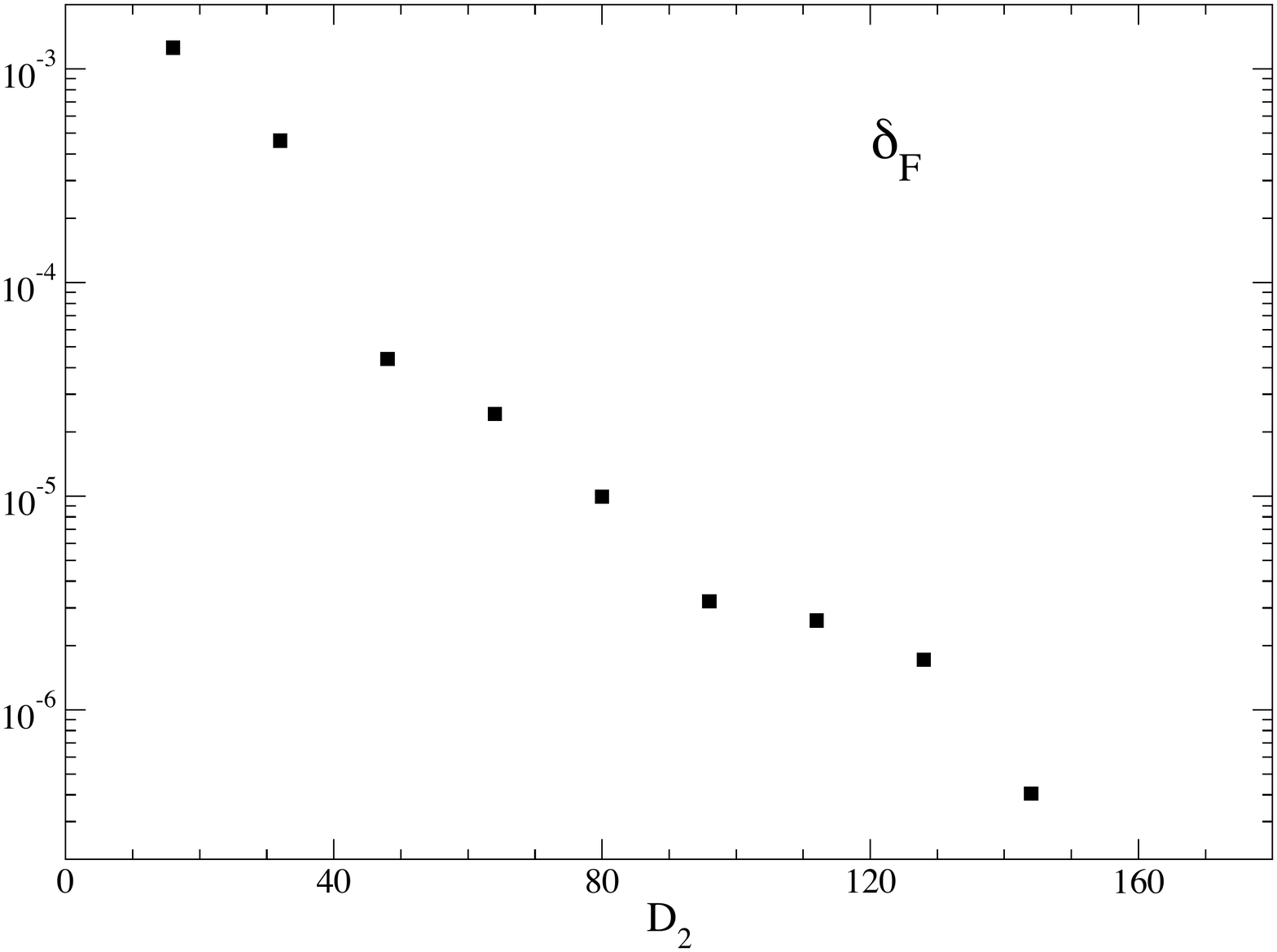}
    \caption{$D_2$ dependence of $(\ln Z) /V$ (top) and  $\delta_F$ (bottom) at $\beta=0.71115$ on the $4096^2\times 3$ lattice.}
    \label{fig:err}
  \end{figure}

Numerical study of 3D Z$_2$ gauge theory is performed by employing the algorithm explained in Sec.~\ref{sec:TNscheme} on the $N_{\sigma}^2\times N_{\tau}$ lattice with the periodic boundary condition. We fix the temporal extension 
at $N_{\tau}=2,3,5$. The spatial lattice size is varied up to $N_{\sigma}=4096$ in order to make finite size scaling analyses toward the thermodynamic limit.  In Fig.~\ref{fig:err} we plot the $D_2$ dependence of $(\ln Z) /V$ and $\delta_F=\vert 1-\ln Z(D_2)/\ln Z(D_2=160)\vert$ at $\beta=0.71115$ on the $4096^2\times 3$ lattice, which illustrate a convergence behavior of $(\ln Z) /V$ almost on the phase transition point as a representative case.  
We observe that the value of $(\ln Z) /V$ monotonically converges 
as $D_2$ increases. Since we find similar behaviors at other $\beta$ values, 
we take the results with $D_2=128$ as the central values and their errors 
are estimated
by the difference from those with $D_2=144$.

  \begin{figure}[h]
    \centering
    \includegraphics[width=80mm]{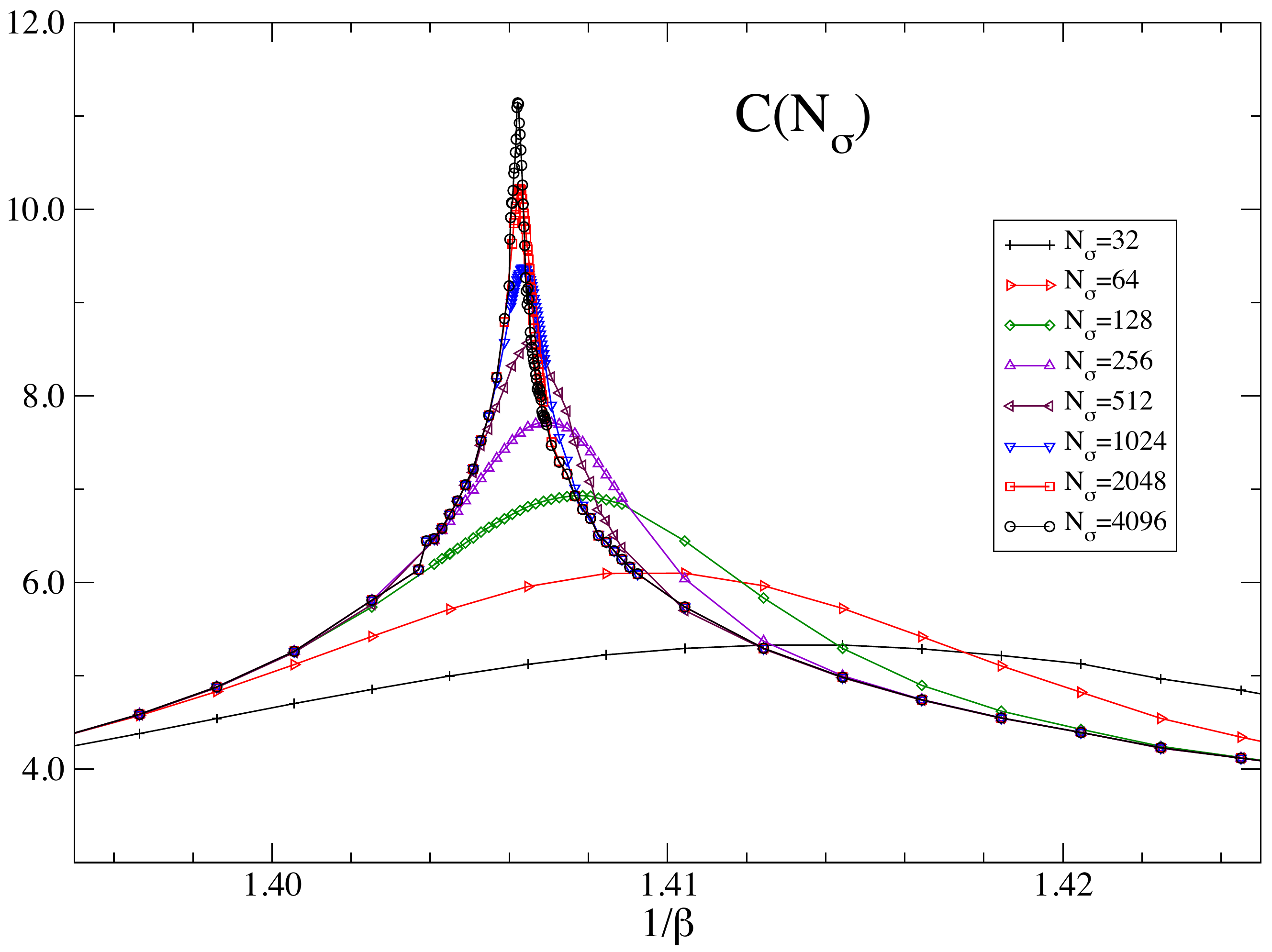}
    \caption{Specific heat $C(N_\sigma)$ at $N_\tau=3$ as a function of $1/\beta$ with $N_\sigma\in[32,4096]$.}
    \label{fig:specificheat}
  \end{figure}

\subsection{Results} 
\label{subsec:results}

In Fig.~\ref{fig:specificheat} we plot the specific heat of Eq.~(\ref{eq:specificheat}) as a function of $1/\beta$. We observe the clear peak structure at all the values of $N_\sigma$ and the peak height $C_{\rm max}(N_{\sigma})$ grows as $N_\sigma$ increases. 
In order to determine the peak position $\beta_c(N_\sigma)$ and 
the peak height $C_{\rm max}(N_{\sigma})$ at each $N_\sigma$
we employ the quadratic approximation of the specific heat around $\beta_c(N_\sigma)$:
\be
C(N_{\sigma})\sim C_{\rm max}(N_{\sigma})+R\left(\frac{1}{\beta}-\frac{1}{\beta_c(N_\sigma)}\right)^2 
\ee
with $R$ a constant.
Instead of fitting the specific heat itself
we fit the internal energy $E$ with the following form around 
$\beta_c(N_\sigma)$: 
\be
E=-\frac{\partial \ln Z}{V\partial \beta}=P + \frac{C_{\rm max}(N_{\sigma})}{\beta} + \frac{R}{3}\left(\frac{1}{\beta}-\frac{1}{\beta_c(N_\sigma)}\right)^3 
\ee
with $P$ another constant.
This procedure suffers from less uncertainties associated with 
the numerical derivative compared to the direct fit of the specific heat itself.
  
\begin{figure}[th]
    \centering
    \includegraphics[width=80mm]{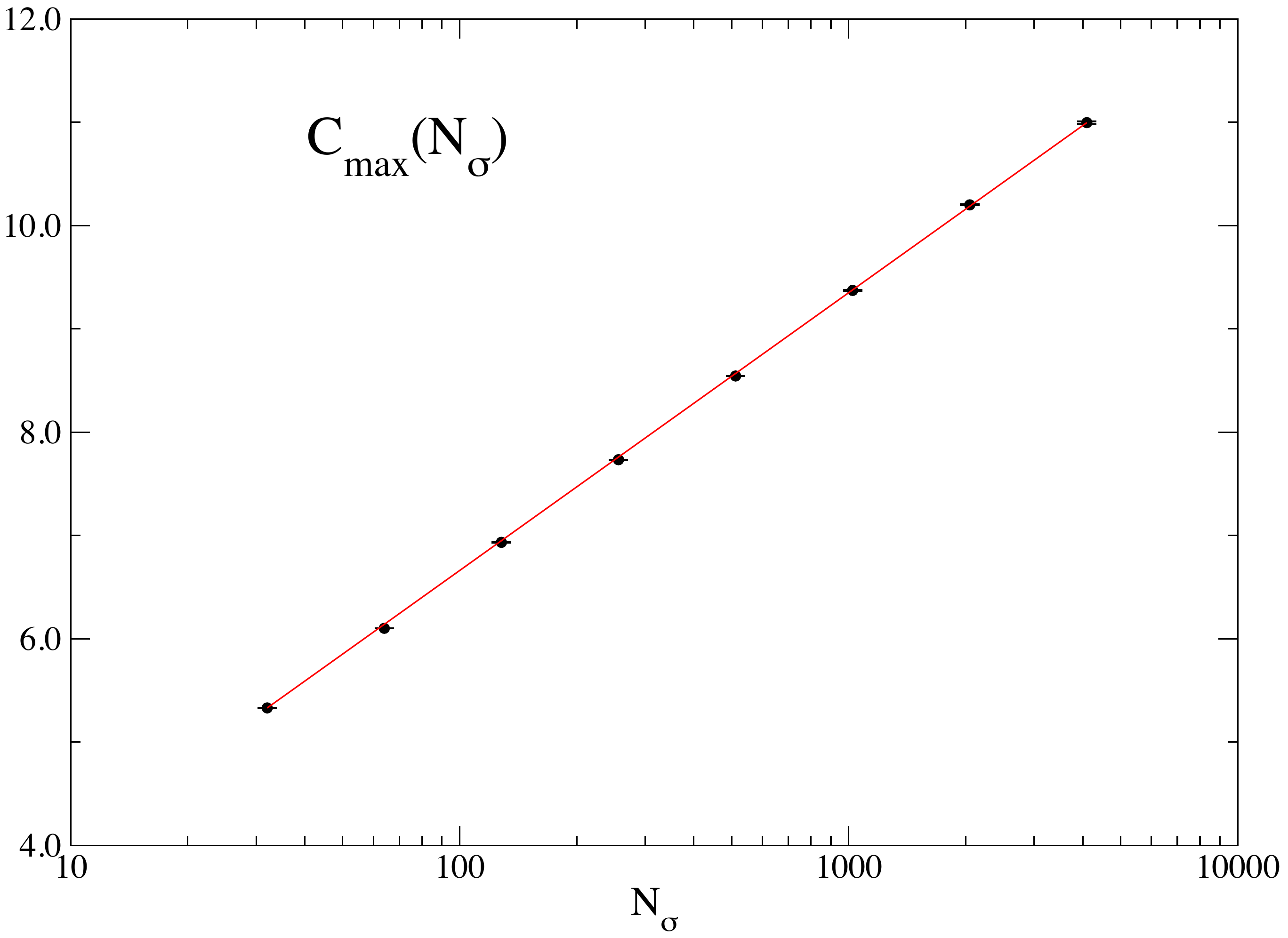}
    \caption{Peak height of specific heat $C_{\rm max}(N_\sigma)$ at $N_\tau=3$ as a function of $N_{\sigma}$. The horizontal axis is logarithmic. Solid line is to guide your eyes.}
    \label{fig:C_max}
  \end{figure}

\begin{figure}[h]
    \centering
    \includegraphics[width=80mm]{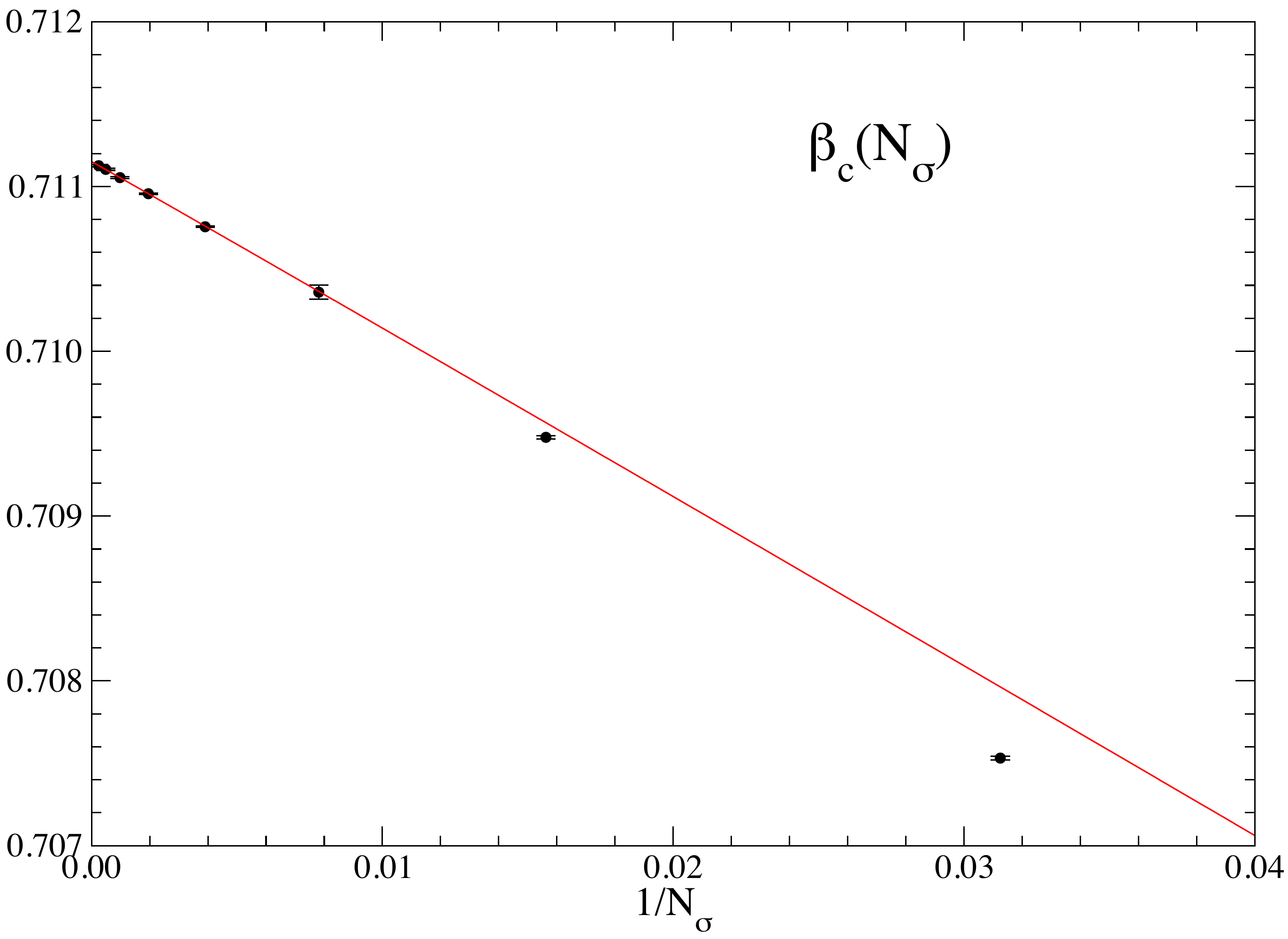}
    \caption{Peak position of the specific heat $\beta_c(N_\sigma)$ at $N_\tau=3$ as a function of $1/N_\sigma$. Solid curve represents the fit result.}
    \label{fig:fss-betac}
  \end{figure}

\begin{figure}[h]
   \centering
   \includegraphics[width=80mm]{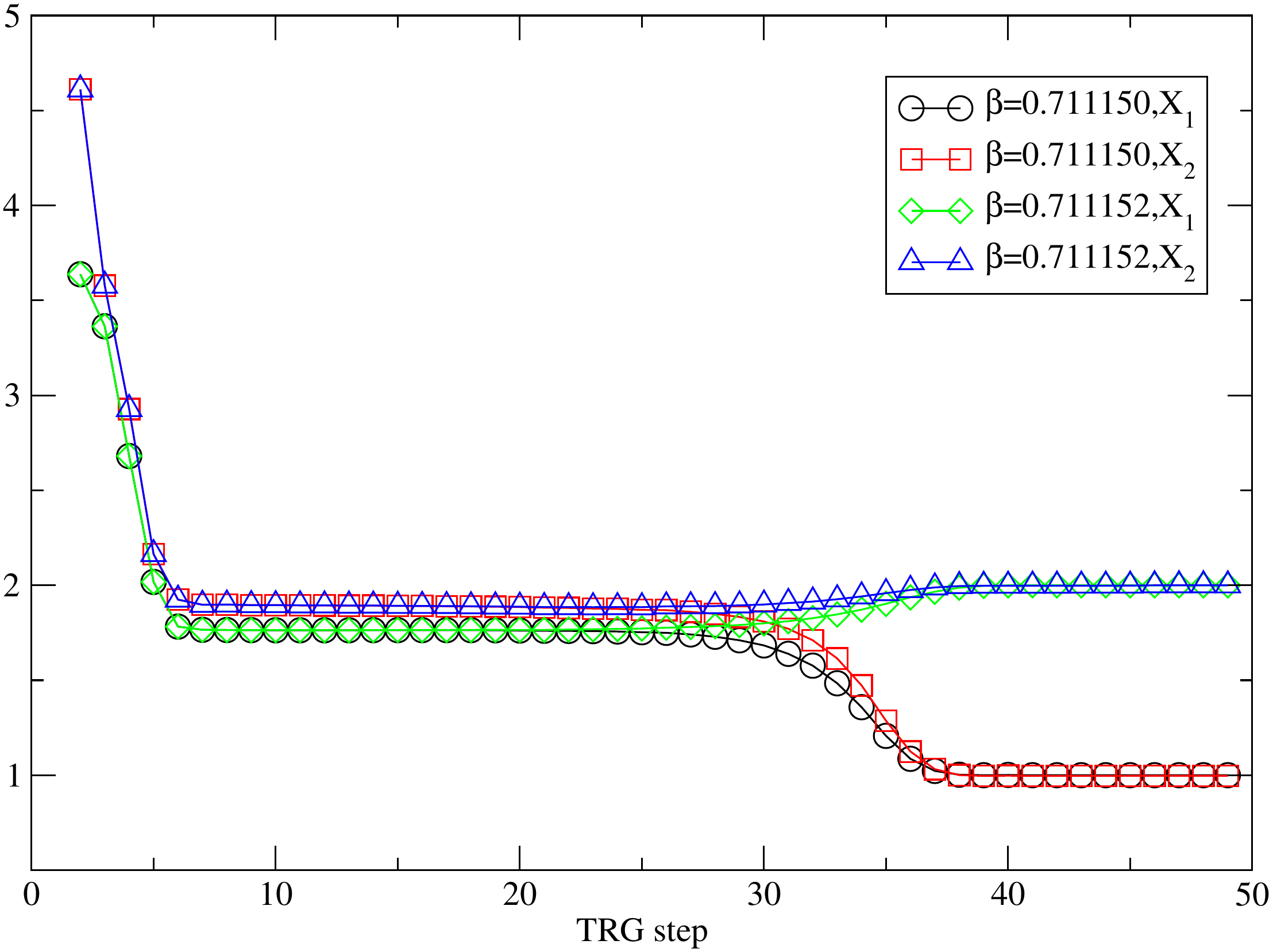}
   \caption{$X_1$ and $X_2$ above and below the transition temperature as a function of TRG steps at $N_\tau=3$.}
   \label{fig:X}
 \end{figure}

We expect that the peak height $C_{\rm max}(N_{\sigma})$ scales with $N_{\sigma}$ as
\ben
C_{\rm max}(N_{\sigma}) \propto N_{\sigma}^{\alpha/\nu},
\een
with the critical exponents $\alpha$ and $\nu$.
We plot the peak height $C_{\rm max}(N_{\sigma})$ at $N_\tau=3$ as a function of $N_{\sigma}$
in Fig.~\ref{fig:C_max}. We observe a clear logarithmic $N_{\sigma}$ dependence for $C_{\rm max}(N_{\sigma})$. 
We have found similar features at other $N_\tau$.
These observation indicates $\alpha\simeq 0$.
We can determine another critical exponent $\nu$ from the finite size scaling behavior of the peak position $\beta_c(N_\sigma)$,
\ben
\beta_c(N_\sigma)-\beta_c(\infty)\propto N_\sigma^{-1/\nu}.
\een
Figure~\ref{fig:fss-betac} shows $N_\sigma$ dependence of $\beta_c(N_\sigma)$ at $N_\tau=3$ as a representative case. The solid curve represents the fit result obtained with the fit function of $\beta_c(N_\sigma)=\beta_c(\infty)+B N_\sigma^{-1/\nu}$. In Table~\ref{tab:betac} we list the fit range at each $N_\tau$ which is chosen to avoid possible finite size effects due to the smaller $N_\sigma$. The fit results for $\beta_c(\infty)$, $B$, $\nu$ and $\chi^2/{\rm d.o.f.}$ are summarized in Table~\ref{tab:betac}. The values of $\beta_c(\infty)$ at $N_\tau=2,3,5$ estimated in Ref.~\cite{z2-mc} are systematically smaller than ours beyond the error bars. This may be attributed to the narrow range of $N_\sigma$ employed in Ref.~\cite{z2-mc}. 
{ In Table~\ref{tab:betac} we also list the transition temperature $\beta_c^{X_1}$ and $\beta_c^{X_2}$ determined with the use of the quantities $X_1$ and $X_2$ introduced in Ref.~\cite{tefr}, which show consistency with $\beta_c(\infty)$ within 2 or 3$\sigma$ error band. Figure~\ref{fig:X} shows a typical behavior of the values of $X_1$ and $X_2$ as a function of TRG steps.}
For the values of $\nu$ we observe that both of our results and those in Ref.~\cite{z2-mc} are consistent with $\nu=1$, which is the expected critical exponent in 2D Ising model. This also satisfies the Josephson law of $d\nu=2-\alpha$ with $d=2$ in the two-dimensional case.
These are supporting evidences for Svetitsky-Yaffe conjecture that the finite temperature transitions in ($d+1$)-dimensional SU($N$) and Z$_N$ lattice gauge theories belong to the same universality class of those in the corresponding $d$-dimensional Z$_N$ spin models \cite{SY-conjecture}. In our case the universality class for the finite temperature Z$_2$ lattice gauge theory should coincide with that for the 2D Ising model whose critical exponents are $\alpha=0$ and $\nu=1$.  

    \begin{table*}[t]
    	\centering
    	\caption{Fit results for the critical point $\beta_c(\infty)$ and the critical exponent $\nu$ at $N_\tau=2,3,5$. $\beta_c^{X_1}$ and $\beta_c^{X_2}$ are determined at $N_\sigma\le 2^{20}$. The value of $\nu$ at $N_\tau=2$ in Ref.~\cite{z2-mc} is evaluated using the pair of data at $N_\sigma=16$ and 32.}
    	\label{tab:betac}
    		\begin{tabular}{cccccccc}
    		\bottomrule
    		& & & This work & & & & \\
    		$N_\tau$ & $N_\sigma$ & $\beta_c(\infty)$ & $\nu$ & $B$ & $\chi^2$/d.o.f. &  $\beta_c^{X_1}$ & $\beta_c^{X_2}$   \\ \hline
    		$2$ & $[512,4096]$ & $0.656097(1)$ & $1.00(1)$ & $0.116(6)$ & $0.086$ & $0.656094(1)$  & $0.656094(1)$ \\
    		$3$ & $[512,4096]$ & $0.711150(4)$ & $0.99(4)$ & $0.10(3)$ & $0.047$  &  $0.711151(1)$ & $0.711151(1)$ \\
    		$5$ & $[512,4096]$ & $0.740730(3)$ & $0.96(5)$ & $0.08(3)$ & $0.012$  & $0.740734(1)$ & $0.740734(1)$ \\ \hline \hline
                         &           & Ref.~\cite{z2-mc} & &&&& \\
     		$N_\tau$ & $N_\sigma$ & $\beta_c(\infty)$ & $\nu$  &&&& \\ \hline 
    		$2$ & $4,8,16,32$ & $0.65608(5)$ & $1.012(21)$ &&&&\\
    		$3$ & $24$ & $0.71102(8)$ & $$ &&&&\\
    		$5$ & $40$ & $0.74057(3)$ & $$ &&&&\\ \hline
		   	\toprule
		   	\end{tabular}
 	\end{table*}
 	
  \section{Summary and outlook}
  \label{sec:concl}
We have applied the tensor network scheme to a study of 3D finite temperature Z$_2$ gauge theory. Its efficiency is demonstrated by a numerical study of the critical properties of the 3D Z$_2$ gauge theory. The tensor network scheme enables us to make a large scale of finite size scaling analysis with the wide range of $N_\sigma$ thanks to the $\ln V$ dependence of the computational cost, which allows us a precise and reliable estimation of the critical point and the critical exponent at the thermodynamic limit.
This is the first successful application of the tensor network scheme to one of the simplest 3D lattice gauge theories. Next step may be the extension of this approach to the gauge theories with continuous groups.  
\vspace*{+2mm}

\begin{acknowledgments}
Numerical calculation for the present work was carried out with the COMA (PACS-IX) computer under the Interdisciplinary Computational Science Program of Center for Computational Sciences, University of Tsukuba.
This work is supported by the Ministry of Education, Culture, Sports, Science and Technology (MEXT) as ``Exploratory Challenge on Post-K computer (Frontiers of Basic Science: Challenging the Limits)'' and also in part by Grants-in-Aid for Scientific Research from MEXT (No. 15H03651).
\end{acknowledgments}


\end{document}